\documentclass[prl,twocolumn,showpacs]{revtex4}
\usepackage{graphicx}
\usepackage{amsmath,amssymb}
\graphicspath{{./img/}}

\newcommand{\tr}{\mathop{\mathrm{tr}}\nolimits}
\renewcommand{\section}[1]{{\par\it #1.---}\ignorespaces}

\newcommand{\void}[1]{}

\begin{document}
\title{Coherent Quantum Ratchets Driven by Tunnel Oscillations}

 \author{Michael Stark}
 \author{Sigmund Kohler}
 \affiliation{Instituto de Ciencia de Materiales de Madrid, CSIC,
	Cantoblanco, 28049 Madrid, Spain}
 \date{\today} 
%
\begin{abstract}
We demonstrate that the tunnel oscillations of a biased
double quantum dot can be employed as driving source for a quantum
ratchet.  As a model, we use two capacitively coupled double quantum
dots.  One double dot is voltage biased and provides the ac force,
while the other experiences the ac force and acts as coherent quantum
ratchet.  The current is obtained from a Bloch-Redfield master
equation which ensures a proper equilibrium limit.  We find
that the two-electron states of the coupled ratchet-drive Hamiltonian
lead to unexpected steps in the ratchet current.
\end{abstract}


\maketitle


The ratchet effect, i.e.\ the induction of a dc current by an ac force
in the absence of any net bias, represents one of the most intriguing
phenomena in the field of non-equilibrium transport
\cite{Reimann2002a, Hanggi2009a}.  Its quantum version
\cite{Reimann1997a}, has been observed, e.g., in nanostructured
two-dimensional electron gases \cite{Linke1999a}, double quantum dots
\cite{vanderWiel2003a}, Josephson junctions \cite{Ustinov2004a,
Sterck2005a}, and Josephson junction arrays \cite{Majer2003a}.  In all
these experiments, spatially asymmetric potentials are driven by an
external ac field stemming from a \textit{classical} radiation source.
By contrast, we here address the question whether the tunnel
oscillations of a single electron can be employed to induce a sizable
ratchet current.

Recently a quantum ratchet has been realized with a double quantum dot
driven by the non-equilibrium noise of a close-by quantum point
contact (``drive circuit'') \cite{Khrapai2006a}.  The observed current
exhibits characteristic ratchet features such as
current reversals and a vanishing current at symmetry points.
If the relevant energy levels of the double quantum dot are strongly
detuned, the inter-dot tunneling is incoherent and occurs at a rate
that can be derived within $P(E)$ theory \cite{Ingold1992a}.
The resulting current is proportional to the noise correlation
function and, thus, the ratchet can serve as frequency-resolved noise
detector \cite{Aguado2000a, Onac2006a, Gustavsson2007a}.  However,
when the detuning becomes of the order of the inter-dot coupling,
coherent tunnel oscillations emerge and, thus, a treatment beyond
$P(E)$ theory becomes necessary.

\begin{figure}[b]
\includegraphics[scale=.9]{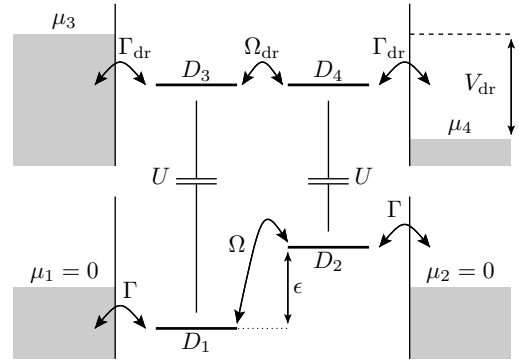}
\caption{Capacitively coupled double quantum dots acting as quantum
ratchet (dots $D_{1,2}$ with onsite energies $\epsilon_1 =
-\epsilon/2$, $\epsilon_2=\epsilon/2$) and drive circuit (dots
$D_{3,4}$ with $\epsilon_3=\epsilon_4=0$), respectively, where each
dot is coupled to a lead.  The tunnel oscillations in the drive
circuit are maintained by a voltage bias, while the ratchet is
detuned, but unbiased.}
\label{fig:setup}
\end{figure}

In the scenario sketched so far, the drive circuit entails a force on
the ratchet, while the corresponding backaction is ignored.
While this is a valid approach for classical driving fields,
it becomes inadequate when the driving force stems from a single
quantum mechanical degree of freedom \cite{note:backaction,
Taubert2008a}.  Therefore, a consistent description requires including
the drive circuit into the model.
We consider the setup sketched in Fig.~\ref{fig:setup}, where both
the ratchet and the drive circuit are formed by double quantum dots
that are capacitively coupled.  Thereby we find even for small capacitive
coupling a sizable ratchet current and, moreover, elucidate the role
of eigenstates with one electron in the drive circuit and one in
the ratchet.

\section{Capacitively coupled double quantum dots}
The setup of Fig.~\ref{fig:setup} is described by the Hamiltonian
$ H=H_\text{dots}+ \sum_\ell V_\ell +\sum_\ell H_\ell$,
where
\begin{equation}
\label{Hd}
\begin{split}
H_\text{dots}
= & \sum_{\ell=1}^4 \epsilon_\ell n_\ell
    - \frac{\Omega}{2}(c_2^\dag c_1+c_1^\dag c_2)
\\ &
    - \frac{\Omega_\text{dr}}{2} (c_4^\dag c_3+c_3^\dag c_4)
    + U ( n_1 n_3 + n_2 n_4 )
\end{split}
\end{equation}
refers to the two double quantum dots.  Each of the four dots
$D_\ell$, $\ell=1,\ldots,4$ is treated as a single level
$|\ell\rangle$ with onsite energy $\epsilon_\ell$.  The operators
$c_\ell^\dag$ and $c_\ell$ create and annihilate, respectively, an
electron on dot $D_\ell$, and $n_\ell = c_\ell^\dag c_\ell$ is the
corresponding number operator.  The second and third terms constitute
electron tunneling between dots $D_1$ and $D_2$ and between $D_3$ and
$D_4$ with tunnel couplings $\Omega$ and $\Omega_\text{dr}$.  The last
term describes the capacitive interaction between neighboring dots of
opposite circuits.  For the present purpose, it is sufficient to
consider only up to one spinless electron per double dot.

Each dot $D_\ell$ is coupled to a lead $\ell$
fully described by $H_\ell
=\sum_{q} \epsilon_{q} c_{\ell q}^\dag c_{\ell q} $ and $\langle
c_{\ell q}^\dag c_{\ell'q'}\rangle = f(\epsilon_q-\mu_\ell)
\delta_{\ell\ell'} \delta_{qq'}$ with chemical potential
$\mu_\ell$ and the Fermi function
$f(x) = [\exp(x/k_BT)+1]^{-1}$.  The dot-lead contact is established
by the tunnel Hamiltonian
$ V_\ell = \sum_q V_{\ell q} c_{\ell q}^\dag c_\ell + \text{h.c.}$
We assume within a wide-band limit
that all coupling strengths $\Gamma_\ell(\epsilon) = 2\pi\sum_q
|V_{\ell q}|^2 \delta(\epsilon-\epsilon_q)$ are energy independent and
that the setup is symmetric such that $\Gamma_1 = \Gamma_2 = \Gamma$
and $\Gamma_3 = \Gamma_4 = \Gamma_\text{dr}$.

By established techniques \cite{Blum1996a}, we derive for the reduced
density operator of the dots, $\rho$, the Bloch-Redfield master
equation (in units with $\hbar=1$)
\begin{equation}
\dot{\rho}
= -i[ H_\text{dots},\rho ]
  -\tr_\text{leads} \int_0^\infty d\tau
   \sum_\ell [V_\ell,[\tilde{V}_\ell(-\tau), R]] ,
\label{bloch-redfield}
\end{equation}
where $R = \rho\otimes\rho_\text{leads}$.  The tilde denotes the
interaction picture operator $\tilde{x}(t) = U_0(t)^\dag x U_0(t)$ with
$U_0(t) = \exp[-i(H_\text{dots}+\sum_\ell H_\ell)t]$ the
propagator in the absence of dot-lead tunneling.
The inter-dot tunneling, however, has to be included in $U_0$ to
ensure compliance with equilibrium conditions \cite{Novotny2002a}.
This is in the present case of particular importance, because
otherwise the master equation would provide a spurious current which
may even be larger than the ratchet current.
The central quantities of interest are the currents through the
dot-lead contacts which we define as the time-derivative of the charge
in the respective lead, $I_\ell = -e(d/dt) N_\ell$.  After some
algebra, we obtain \cite{Kohler2005a} $I_\ell =
\tr(\mathcal{J}_\ell^\text{out} -\mathcal{J}_\ell^\text{in}) \rho$,
where
\begin{equation}
\label{Jout}
\mathcal{J}_\ell^\text{out} \rho
=\frac{e\Gamma_\ell}{2\pi}\int_0^\infty d\tau\int d\epsilon\,
  e^{-i\epsilon\tau}\tilde{c}_\ell(-\tau)
  \rho c_\ell^\dag f_\ell(\epsilon) +\text{h.c.},
\end{equation}
and $\mathcal{J}_\ell^\text{in}$ is formally
obtained from $\mathcal{J}_\ell^\text{out}$ by the replacement
$(c_\ell^\dagger,c_\ell,f_\ell) \to (c_\ell,c_\ell^\dagger,
1-f_\ell)$.

For the numerical solution of the master equation
\eqref{bloch-redfield}, we need to cope with the interaction picture
representation of the tunneling operators $V_\ell$.  For the lead
operators, we readily insert $c_{\ell q}(t) = c_{\ell
q} \exp(-i\epsilon_q t)$, while for the dot operators, we accomplish
this task by decomposing both the master equation and the current
operators into the eigenstates of $H_\text{dots}$.
Thus, we have to solve the eigenvalue equation $H_\text{dots}
|\phi_\alpha^{(n)}\rangle = E_\alpha^{(n)} |\phi_\alpha^{(n)}\rangle$.
The complementary index $(n)$ reflects the respective electron number
$n \equiv n_\alpha =\langle\phi_\alpha^{(n)}|\sum_\ell
n_\ell |\phi_\alpha^{(n)}\rangle$.
Then the master equation assumes the form $\dot\rho_{\alpha\beta} = -i
[E_\alpha^{(n)}-E_\beta^{(n)}] \rho_{\alpha\beta}
+\sum_{\alpha'\beta'} \mathcal{L}_{\alpha\beta,\alpha'\beta'}
\rho_{\alpha'\beta'}$.
The full expression for $\mathcal{L}_{\alpha\beta,\alpha'\beta'}$ is
somewhat lengthy so that we do not write it explicitly.

\section{Stochastic ac driving}
Before addressing the question how the ratchet acts back on the drive
circuit, we work out the scenario in which electrons tunneling through
the drive circuit entail an effective ac force on the ratchet.  In
doing so, we generalize the previous $P(E)$ theory treatment
\cite{Aguado2000a} to the case of delocalized ratchet electrons.

The effective ac driving can be obtained from Hamiltonian
\eqref{Hd} as follows.  An electron on dot $D_{3}$ shifts the onsite
energy $\epsilon_{1}$ by $U$, while $\epsilon_2$ is shifted by $U$ if
an electron resides on dot $D_4$, i.e.\ the ratchet detuning
$\epsilon = \epsilon_2-\epsilon_1$ changes by $U\xi$,
where $\xi = n_4-n_3$.  Hence the ratchet acquires the stochastic
Hamiltonian $H_\text{noise} = \frac{U}{2}\xi(n_2-n_1)$.  For its
treatment with Fermi's golden rule, we need to compute the Fourier
transformed $\hat C(\omega)$ of the correlation function $C(t) =
\langle \xi(t) \xi(0)\rangle$.  For large bias voltage $\mu_3-\mu_4
{} \gg \Omega_\text{dr}$, all levels of the drive circuit lie within the
voltage window.  Then the corresponding Bloch-Redfield equation
assumes the Lindblad form $\mathcal{L}_\text{dr} =
-\frac{i}{2}\Omega_\text{dr}[c_3^\dag c_4+c_4^\dag c_3,\rho] +
\Gamma_\text{dr}\mathcal{D}(c_3^\dag)\rho +
\Gamma_\text{dr}\mathcal{D}(c_4)\rho$ with the superoperator
$\mathcal{D}(x)\rho = x\rho x^\dag - \frac{1}{2}x^\dag x\rho -
\frac{1}{2}\rho x^\dag x$.  Employing the quantum regression
theorem \cite{Lax1963a}, we obtain
\begin{equation}
\label{Cdrive}
\hat C(\omega)
= \frac{\Gamma_\text{dr}^2+2\Omega_\text{dr}^2}
       {\Gamma_\text{dr}^2+3\Omega_\text{dr}^2}
  \frac{\Gamma_\text{dr}/2}{(\omega-\Omega_\text{dr})^2+\Gamma_\text{dr}^2/4}
  + (\Omega_\text{dr}\to-\Omega_\text{dr}) ,
\end{equation}
i.e.\ a double Lorentzian with peaks at $\pm\Omega_\text{dr}$.
In the time domain, $C(t) \propto \cos(\Omega_\text{dr}t)
\exp(-\Gamma_\text{dr}t/2)$ which underlines the ac character of the
stochastic force $\xi$ for $\Gamma_\text{dr} \ll \Omega_\text{dr}$.

\begin{figure}[bt]
\includegraphics[scale=.9]{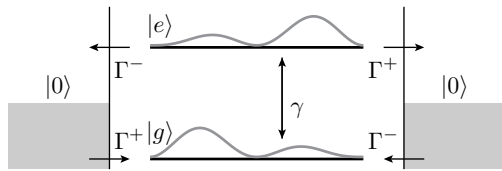}
\caption{Transitions between the one-electron states $|g\rangle$,
$|e\rangle$ and the empty state $|0\rangle$ of the stochastically ac
driven ratchet.
\label{fig:rates}
}
\end{figure}

The Hamiltonian $H_\text{noise}$ induces transitions between the
one-electron eigenstates of the ratchet Hamiltonian, $|g\rangle =
\cos\theta|1\rangle +\sin\theta|2\rangle$ and $|e\rangle =
-\sin\theta|1\rangle +\cos\theta|2\rangle$ with $\cos(2\theta) =
\epsilon/E$ and the level splitting $E=(\epsilon^2+\Omega^2)^{1/2}$.
For sufficiently small $U$, a golden rule calculation with the
transition matrix element $\langle e|H_\text{noise}|g\rangle =
\frac{1}{2}U\xi\sin(2\theta)$ yields the rate $\gamma =
\frac{1}{4} U^2 \sin^2(2\theta) \hat C(E)$.
Once the electron is in the excited state $|e\rangle$, it may tunnel
to lead~1 or to lead~2.  In turn, if the ratchet double dot is in the
empty state $|0\rangle$, an electron may tunnel from one of the leads
to the ground state $|g\rangle$.  By transforming the current
operators $\mathcal{J}_{1,2}^\text{out/in}$ into the basis
$\{|e\rangle,|g\rangle\}$, we find that the transition rates are
proportional to the overlaps $|\langle \ell|e\rangle|^2$, $\ell=1,2$.
Thus we obtain for transitions between the ratchet states the
stochastic process sketched in Fig.~\ref{fig:rates} with the rates
$\Gamma^+ = \Gamma\cos^2\theta$ and $\Gamma^- = \Gamma\sin^2\theta$.

It is straightforward \cite{vanKampen1992a} to find for the occupation
probabilities the master equation
\begin{equation}
\frac{d}{dt}
\begin{pmatrix} p_0 \\ p_g \\ p_e \end{pmatrix}
=
\begin{pmatrix} -\Gamma & 0 & \Gamma \\
               \Gamma & -\gamma & \gamma \\
                 0 & \gamma & -(\Gamma+\gamma)
\end{pmatrix}
\begin{pmatrix} p_0 \\ p_g \\ p_e \end{pmatrix} ,
\end{equation}
and the current $I = e(\Gamma^+p_e -\Gamma^-p_0)$, where
$\Gamma=\Gamma^++\Gamma^-$.  From the stationary solution
$(p_0,p_g,p_e) \propto
(\gamma, \Gamma+\gamma, \gamma)$ follows $I = e
\gamma(\Gamma^+-\Gamma^-)/(\Gamma^++\Gamma^-+3\gamma)$.
We insert the above expressions for $\gamma$ and $\Gamma^{\pm}$ and
express the mixing angle $\theta$ in terms of $\epsilon$ and $\Omega$
to obtain in the limit $\gamma\ll\Gamma$ the ratchet current
\begin{equation}
\label{Igr}
I
= \frac{eU^2}{4}\frac{\epsilon\Omega^2}{(\epsilon^2+\Omega^2)^{3/2}}
  \hat C\Big(\sqrt{\epsilon^2+\Omega^2}\Big)
.
\end{equation}
Notice that the second factor is beyond $P(E)$ theory and represents
the essential difference to Ref.~\cite{Aguado2000a}.  Its origin is
the delocalization of the ratchet eigenstates for
$\epsilon\lesssim\Omega$.  In the present context, this regime is the
most intriguing one, because it contains both the current maximum at
$\epsilon\approx \Omega$ and the main current reversal at
$\epsilon=0$.

The analytical result \eqref{Igr}
already allows an estimate for the size of the ratchet
current.  The Lorentzian $\hat C$ assumes its maximum
$\sim1/\Gamma_\text{dr}$ if the ratchet and the drive circuit are in
resonance, $E = \Omega_\text{dr}$.  The corresponding condition on the
second factor of this expression is $\epsilon = \Omega/\sqrt{2}$, so
that the maximal ratchet current is roughly $I_\text{max} =
eU^2/5\Gamma_\text{dr}$.  Interestingly enough, this value depends
only on the parameters of the drive circuit and on the coupling
strength, but not on the ratchet parameters.

In the experiment of Ref.~\cite{Khrapai2006a}, the dot-lead coupling
is $40\mu\mathrm{eV}$, while the capacitive coupling $U$ is
significantly smaller.  Assuming $U=0.2\mu\mathrm{eV}$, we obtain
$I_\text{max} \approx 0.1\mathrm{pA}$, i.e.\ the appreciable value
measured for driving with a quantum point contact \cite{Khrapai2006a}.
Two double quantum dots with similar geometry and similar tunnel
couplings but with the much stronger interaction $U =
20\mu\mathrm{eV}$ have already been realized \cite{Petersson2009a},
such that a considerably larger ratchet current should be achievable
as well.

\section{Backaction on the drive circuit}
\begin{figure}[bt]
\includegraphics[scale=.9]{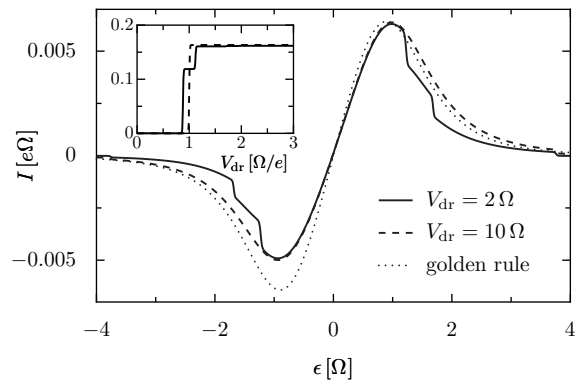}
\caption{Ratchet current as a function of the level detuning for
various bias voltages at the drive circuit in comparison with the
golden-rule result \eqref{Igr}.  The tunnel couplings are $\Gamma =
0.3\,\Omega$, $\Gamma_\text{dr} = 0.5\,\Omega$, and $\Omega_\text{dr} =
\Omega$, while the interaction strength is $U = 0.25\,\Omega$.
Inset: drive current as a function of $V_\text{dr}$ for $U=0$ (dashed)
and $U=0.25\Omega$ (solid), while $\epsilon = 2\Omega$.}
\label{fig:I(e)}
\end{figure}

Let us now turn to the treatment of the drive circuit as a quantum
system that is affected by the coupling to the ratchet.  For this
purpose, we compute the currents numerically by solving the master
equation \eqref{bloch-redfield}.
Figure~\ref{fig:I(e)} demonstrates that for the large
drive circuit bias $V_\text{dr} = 10\Omega/e$, the ratchet current
agrees quite well with our prediction~\eqref{Igr}.  In particular, it
exhibits a current reversal at $\epsilon=0$, while the current maximum
is obtained for $\epsilon \approx \pm\Omega$, i.e.\ in the coherent
regime.  The lack of perfect symmetry is due to the fact that on
average, dot $D_3$ is slightly stronger populated than dot $D_4$.  The
drive current (inset of Fig.~\ref{fig:I(e)}) is influenced by the
interaction only close to $V_\text{dr} \approx \Omega_\text{dr}/e$.
For larger $V_\text{dr}$, the drive current stays practically
constant, so that in the picture of stochastic ac driving, no change
of the ratchet current is expected.

In contrast to this expectation,
however, e.g.\ for $V_\text{dr} =2\Omega/e$ (Fig.~\ref{fig:I(e)}),
the ratchet current exhibits steps similar to those of Coulomb
blockade.  The location of these steps is best visible in the
differential transconductance $\partial I(\epsilon,
V_\text{dr})/\partial V_\text{dr}$ shown in Fig.~\ref{fig:I(e,V)}(b).
They are based on the fact that transitions between states with different
electron number require the corresponding energy difference to lie
within the voltage window.  Thus, we can identify the states that
govern the transport yielding a full quantum mechanical picture of the
ratchet mechanism.  By investigating the location of the steps
upon variation of $\mu_3$ and $\mu_4$, we find that they relate to the
transitions marked in Fig.~\ref{fig:I(e,V)}(a).  All relevant
transitions involve two-electron states, since the ratchet
current is interaction induced.
\begin{figure}[bt]
\includegraphics[scale=.9]{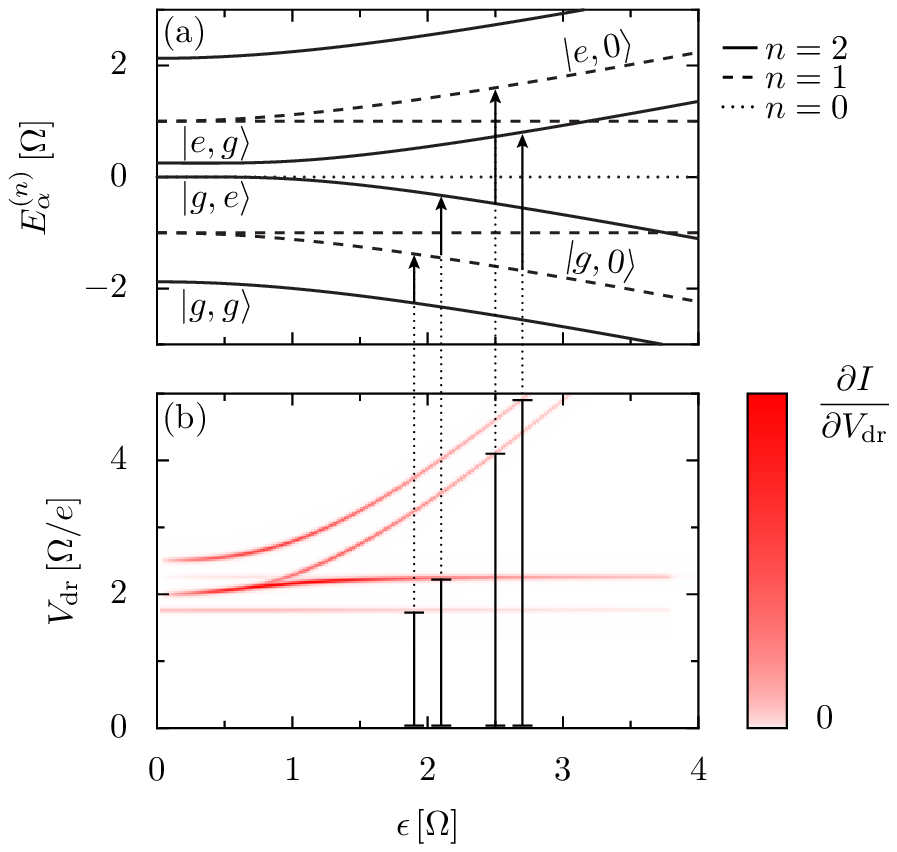}
\caption{(a) Eigenenergies of the $n$-electron states
$|\phi_\alpha^{(n)}\rangle$ of the couped double quantum dots for
$U=0.25\,\Omega$.  The arrows mark the transitions that govern the
ratchet current.
(b) Differential transconductance $\partial I/\partial V_\text{dr}$
highlighting the steps of the ratchet current as a function of the
detuning and the bias at the drive circuit.
All other parameters are as in Fig.~\ref{fig:I(e)}.}
\label{fig:I(e,V)}
\end{figure}

A more profound discussion of the transport process requires knowledge
about the structure of the eigenstates of Hamiltonian \eqref{Hd}.  For
$U=0$, they are given by the direct products of the ratchet
states $|0\rangle$, $|g\rangle$, $|e\rangle$ and the according
eigenstates of the drive dots, $|0\rangle_\text{dr}$,
$|g\rangle_\text{dr}$, and $|e\rangle_\text{dr}$.
Thus, the one-electron states read $|s,0\rangle =
|s\rangle|0\rangle_\text{dr}$ and $|0,s\rangle =
|0\rangle|s\rangle_\text{dr}$, while the two-electron states are
$|s,s'\rangle = |s\rangle|s'\rangle_\text{dr}$, where
$s,s'\in\{g,e\}$.
Obviously, the interaction does not affect the one-electron states,
while any two-electron state acquires for finite $U$ an admixture of
all other two-electron states.  From standard perturbation theory
follows that the admixture is of the order $U^2$.  For the small
values of $U$ considered here, the admixture is small as well and,
thus, it is appropriate to keep the notation $|s,s'\rangle$.
The consequence of the interaction is that when an electron tunnels from
lead~3 via the drive dots to lead~4, the ratchet ground state gains
a contribution of the excited ratchet state, i.e.\ the drive current
induces a transition of the type $|g\rangle \to |g\rangle +
\lambda|e\rangle$, where $\lambda = \mathcal{O}(U^2)$.  Since an
excited ratchet electron will leave the dots predominantly towards a
particular lead (for $\epsilon>0$ to lead~2), a ratchet current
$I\propto U^2$ will flow.

With this general scenario in mind, it is possible to explain how upon
increasing $V_\text{dr}$, the particular tunnel events come into play.
For zero bias voltage, both the ratchet and the drive circuit are at
equilibrium in state $|g,g\rangle$.  As soon as $\mu_4 =
-V_\text{dr}/2$ drops below the ground
state energy of the drive circuit, both a drive current and
a tiny ratchet current set in.  A second step in the ratchet
current is observed when both drive states lie within the voltage
window and can be occupied.  Only then an electron in the drive
circuit performs coherent tunnel oscillations and entail a
significant ac force.

The third step relates to the transition $|g,e\rangle \to
|e,0\rangle$, i.e.\ an electron tunneling from the drive dots to lead~4
while transferring approximately the energy $\epsilon$ to the ratchet and
thereby exciting it.  Energy conservation now requires that the drive
electron finds in lead~4 an unoccupied state with energy $-\epsilon$,
which is the case for $V_\text{dr} \gtrsim \epsilon$.
Shortly after that, tunneling from lead~3 to the drive dot under
ratchet excitation, i.e.\ the transition $|g,0\rangle \to |e,g\rangle$
is enabled as well, causing the forth current step.
The latter two tunnel processes have a relatively low probability, so
that their influence on the drive current is not noticeable.  By
contrast, their impact on the ratchet current is significant, because
they immediately lead to the directed transport of an ratchet
electron.
Since this scenario relies on the interaction-induced high-energy
components of the drive circuit states, the full ratchet mechanism is
active only at unexpectedly large voltages.  The underlying formation
of two-electron states is not included in the picture of stochastic ac
driving and, thus, represents the relevant backaction.

\section{Conclusions}
We have studied a coherent double-dot quantum ratchet
similar to that of recent experiments \cite{Khrapai2006a,
Gustavsson2007a}, but with a driving stemming from tunnel
oscillations in a close-by further double quantum dot.  These tunnel
oscillations turned out to be sufficiently strong and stable to induce
ratchet currents of the order of those in related experiments.
A particular feature of our model is that it includes both the ratchet
and the drive circuit, which enabled us to go beyond the picture in
which the drive circuit is not affected by the ratchet.  This revealed
that the states of the drive circuit acquire components with high
energies, which represents the relevant backaction by the ratchet.
The measurable consequences are unexpected steps in the ratchet
current.
Moreover, our results imply that tunnel oscillations can be employed
as on-chip sources of ac driving.  Thus we are confident that our
results will stimulate further experimental theoretical effort in this
direction.

We thank G. Platero and F. Dominguez for helpful discussions.
M.S. acknowledges funding by DAAD.
S.K. is supported by a Ram\'on y Cajal fellowship.




\begin{thebibliography}{21}
\expandafter\ifx\csname natexlab\endcsname\relax\def\natexlab#1{#1}\fi
\expandafter\ifx\csname bibnamefont\endcsname\relax
  \def\bibnamefont#1{#1}\fi
\expandafter\ifx\csname bibfnamefont\endcsname\relax
  \def\bibfnamefont#1{#1}\fi
\expandafter\ifx\csname citenamefont\endcsname\relax
  \def\citenamefont#1{#1}\fi
\expandafter\ifx\csname url\endcsname\relax
  \def\url#1{\texttt{#1}}\fi
\expandafter\ifx\csname urlprefix\endcsname\relax\def\urlprefix{URL }\fi
\providecommand{\bibinfo}[2]{#2}
\providecommand{\eprint}[2][]{\url{#2}}

\bibitem[{\citenamefont{Reimann}(2002)}]{Reimann2002a}
\bibinfo{author}{\bibfnamefont{P.}~\bibnamefont{Reimann}},
  \bibinfo{journal}{Phys. Rep.} \textbf{\bibinfo{volume}{361}},
  \bibinfo{pages}{57} (\bibinfo{year}{2002}).

\bibitem[{\citenamefont{H\"anggi and Marchesoni}(2009)}]{Hanggi2009a}
\bibinfo{author}{\bibfnamefont{P.}~\bibnamefont{H\"anggi}} \bibnamefont{and}
  \bibinfo{author}{\bibfnamefont{F.}~\bibnamefont{Marchesoni}},
  \bibinfo{journal}{Rev. Mod. Phys.} \textbf{\bibinfo{volume}{81}},
  \bibinfo{pages}{387} (\bibinfo{year}{2009}).

\bibitem[{\citenamefont{Reimann et~al.}(1997)\citenamefont{Reimann, Grifoni,
  and H\"anggi}}]{Reimann1997a}
\bibinfo{author}{\bibfnamefont{P.}~\bibnamefont{Reimann}},
  \bibinfo{author}{\bibfnamefont{M.}~\bibnamefont{Grifoni}}, \bibnamefont{and}
  \bibinfo{author}{\bibfnamefont{P.}~\bibnamefont{H\"anggi}},
  \bibinfo{journal}{Phys. Rev. Lett.} \textbf{\bibinfo{volume}{79}},
  \bibinfo{pages}{10} (\bibinfo{year}{1997}).

\bibitem[{\citenamefont{Linke et~al.}(1999)\citenamefont{Linke, Humphrey,
  L\"ofgren, Shuskov, Newbury, Taylor, and Omling}}]{Linke1999a}
\bibinfo{author}{\bibfnamefont{H.}~\bibnamefont{Linke}},
  \bibinfo{author}{\bibfnamefont{T.~E.} \bibnamefont{Humphrey}},
  \bibinfo{author}{\bibfnamefont{A.}~\bibnamefont{L\"ofgren}},
  \bibinfo{author}{\bibfnamefont{A.~O.} \bibnamefont{Shuskov}},
  \bibinfo{author}{\bibfnamefont{R.}~\bibnamefont{Newbury}},
  \bibinfo{author}{\bibfnamefont{R.~P.} \bibnamefont{Taylor}},
  \bibnamefont{and} \bibinfo{author}{\bibfnamefont{P.}~\bibnamefont{Omling}},
  \bibinfo{journal}{Science} \textbf{\bibinfo{volume}{286}},
  \bibinfo{pages}{2314} (\bibinfo{year}{1999}).

\bibitem[{\citenamefont{van~der Wiel et~al.}(2003)\citenamefont{van~der Wiel,
  {D}e Franceschi, Elzerman, Fujisawa, Tarucha, and
  Kouwenhoven}}]{vanderWiel2003a}
\bibinfo{author}{\bibfnamefont{W.~G.} \bibnamefont{van~der Wiel}},
  \bibinfo{author}{\bibfnamefont{S.}~\bibnamefont{{D}e Franceschi}},
  \bibinfo{author}{\bibfnamefont{J.~M.} \bibnamefont{Elzerman}},
  \bibinfo{author}{\bibfnamefont{T.}~\bibnamefont{Fujisawa}},
  \bibinfo{author}{\bibfnamefont{S.}~\bibnamefont{Tarucha}}, \bibnamefont{and}
  \bibinfo{author}{\bibfnamefont{L.~P.} \bibnamefont{Kouwenhoven}},
  \bibinfo{journal}{Rev. Mod. Phys.} \textbf{\bibinfo{volume}{75}},
  \bibinfo{pages}{1} (\bibinfo{year}{2003}).

\bibitem[{\citenamefont{Ustinov et~al.}(2004)\citenamefont{Ustinov, Coqui,
  Kemp, Zolotaryuk, and Salerno}}]{Ustinov2004a}
\bibinfo{author}{\bibfnamefont{A.~V.} \bibnamefont{Ustinov}},
  \bibinfo{author}{\bibfnamefont{C.}~\bibnamefont{Coqui}},
  \bibinfo{author}{\bibfnamefont{A.}~\bibnamefont{Kemp}},
  \bibinfo{author}{\bibfnamefont{Y.}~\bibnamefont{Zolotaryuk}},
  \bibnamefont{and} \bibinfo{author}{\bibfnamefont{M.}~\bibnamefont{Salerno}},
  \bibinfo{journal}{Phys. Rev. Lett.} \textbf{\bibinfo{volume}{93}},
  \bibinfo{pages}{087001} (\bibinfo{year}{2004}).

\bibitem[{\citenamefont{Sterck et~al.}(2005)\citenamefont{Sterck, Kleiner, and
  Koelle}}]{Sterck2005a}
\bibinfo{author}{\bibfnamefont{A.}~\bibnamefont{Sterck}},
  \bibinfo{author}{\bibfnamefont{R.}~\bibnamefont{Kleiner}}, \bibnamefont{and}
  \bibinfo{author}{\bibfnamefont{D.}~\bibnamefont{Koelle}},
  \bibinfo{journal}{Phys. Rev. Lett.} \textbf{\bibinfo{volume}{95}},
  \bibinfo{pages}{177006} (\bibinfo{year}{2005}).

\bibitem[{\citenamefont{Majer et~al.}(2003)\citenamefont{Majer, Peguiron,
  Grifoni, Tusveld, and Mooij}}]{Majer2003a}
\bibinfo{author}{\bibfnamefont{J.~B.} \bibnamefont{Majer}},
  \bibinfo{author}{\bibfnamefont{J.}~\bibnamefont{Peguiron}},
  \bibinfo{author}{\bibfnamefont{M.}~\bibnamefont{Grifoni}},
  \bibinfo{author}{\bibfnamefont{M.}~\bibnamefont{Tusveld}}, \bibnamefont{and}
  \bibinfo{author}{\bibfnamefont{J.~E.} \bibnamefont{Mooij}},
  \bibinfo{journal}{Phys. Rev. Lett.} \textbf{\bibinfo{volume}{90}},
  \bibinfo{pages}{056802} (\bibinfo{year}{2003}).

\bibitem[{\citenamefont{Khrapai et~al.}(2006)\citenamefont{Khrapai, Ludwig,
  Kotthaus, Tranitz, and Wegscheider}}]{Khrapai2006a}
\bibinfo{author}{\bibfnamefont{V.~S.} \bibnamefont{Khrapai}},
  \bibinfo{author}{\bibfnamefont{S.}~\bibnamefont{Ludwig}},
  \bibinfo{author}{\bibfnamefont{J.~P.} \bibnamefont{Kotthaus}},
  \bibinfo{author}{\bibfnamefont{H.~P.} \bibnamefont{Tranitz}},
  \bibnamefont{and}
  \bibinfo{author}{\bibfnamefont{W.}~\bibnamefont{Wegscheider}},
  \bibinfo{journal}{Phys. Rev. Lett.} \textbf{\bibinfo{volume}{97}},
  \bibinfo{pages}{176803} (\bibinfo{year}{2006}).

\bibitem[{\citenamefont{Ingold and {Y}u. V.~Nazarov}(1992)}]{Ingold1992a}
\bibinfo{author}{\bibfnamefont{G.-L.} \bibnamefont{Ingold}} \bibnamefont{and}
  \bibinfo{author}{\bibnamefont{{Y}u. V.~Nazarov}},
  \emph{\bibinfo{title}{Charge Tunneling Rates in Ultrasmall Junctions}}
  (\bibinfo{publisher}{Plenum}, \bibinfo{address}{New York},
  \bibinfo{year}{1992}), vol. \bibinfo{volume}{294} of
  \emph{\bibinfo{series}{NATO ASI Series B}}, pp. \bibinfo{pages}{21--107}.

\bibitem[{\citenamefont{Aguado and Kouwenhoven}(2000)}]{Aguado2000a}
\bibinfo{author}{\bibfnamefont{R.}~\bibnamefont{Aguado}} \bibnamefont{and}
  \bibinfo{author}{\bibfnamefont{L.~P.} \bibnamefont{Kouwenhoven}},
  \bibinfo{journal}{Phys. Rev. Lett.} \textbf{\bibinfo{volume}{84}},
  \bibinfo{pages}{1986} (\bibinfo{year}{2000}).

\bibitem[{\citenamefont{Onac et~al.}(2006)\citenamefont{Onac, Balestro, van
  Beveren, Hartmann, Nazarov, and Kouwenhoven}}]{Onac2006a}
\bibinfo{author}{\bibfnamefont{E.}~\bibnamefont{Onac}},
  \bibinfo{author}{\bibfnamefont{F.}~\bibnamefont{Balestro}},
  \bibinfo{author}{\bibfnamefont{L.~H.~W.} \bibnamefont{van Beveren}},
  \bibinfo{author}{\bibfnamefont{U.}~\bibnamefont{Hartmann}},
  \bibinfo{author}{\bibfnamefont{Y.~V.} \bibnamefont{Nazarov}},
  \bibnamefont{and} \bibinfo{author}{\bibfnamefont{L.~P.}
  \bibnamefont{Kouwenhoven}}, \bibinfo{journal}{Phys. Rev. Lett.}
  \textbf{\bibinfo{volume}{96}}, \bibinfo{pages}{176601}
  (\bibinfo{year}{2006}).

\bibitem[{\citenamefont{Gustavsson et~al.}(2007)\citenamefont{Gustavsson,
  Studer, Leturcq, Ihn, Ensslin, Driscoll, and Gossard}}]{Gustavsson2007a}
\bibinfo{author}{\bibfnamefont{S.}~\bibnamefont{Gustavsson}},
  \bibinfo{author}{\bibfnamefont{M.}~\bibnamefont{Studer}},
  \bibinfo{author}{\bibfnamefont{R.}~\bibnamefont{Leturcq}},
  \bibinfo{author}{\bibfnamefont{T.}~\bibnamefont{Ihn}},
  \bibinfo{author}{\bibfnamefont{K.}~\bibnamefont{Ensslin}},
  \bibinfo{author}{\bibfnamefont{D.~C.} \bibnamefont{Driscoll}},
  \bibnamefont{and} \bibinfo{author}{\bibfnamefont{A.~C.}
  \bibnamefont{Gossard}}, \bibinfo{journal}{Phys. Rev. Lett.}
  \textbf{\bibinfo{volume}{99}}, \bibinfo{pages}{206804}
  (\bibinfo{year}{2007}).

\bibitem[{not()}]{note:backaction}
\bibinfo{note}{Measurements of the drive current in the setup of
  Ref.~\cite{Khrapai2006a} indicate a relevant backaction of the ratchet to the
  quantum point contact \cite{Taubert2008a}}.

\bibitem[{\citenamefont{Taubert et~al.}(2008)\citenamefont{Taubert,
  Pioro-Ladri\`ere, Schr\"oer, Harbusch, Sachrajda, and Ludwig}}]{Taubert2008a}
\bibinfo{author}{\bibfnamefont{D.}~\bibnamefont{Taubert}},
  \bibinfo{author}{\bibfnamefont{M.}~\bibnamefont{Pioro-Ladri\`ere}},
  \bibinfo{author}{\bibfnamefont{D.}~\bibnamefont{Schr\"oer}},
  \bibinfo{author}{\bibfnamefont{D.}~\bibnamefont{Harbusch}},
  \bibinfo{author}{\bibfnamefont{A.~S.} \bibnamefont{Sachrajda}},
  \bibnamefont{and} \bibinfo{author}{\bibfnamefont{S.}~\bibnamefont{Ludwig}},
  \bibinfo{journal}{Phys. Rev. Lett.} \textbf{\bibinfo{volume}{100}},
  \bibinfo{pages}{176805} (\bibinfo{year}{2008}).

\bibitem[{\citenamefont{Blum}(1996)}]{Blum1996a}
\bibinfo{author}{\bibfnamefont{K.}~\bibnamefont{Blum}},
  \emph{\bibinfo{title}{Density Matrix Theory and Applications}}
  (\bibinfo{publisher}{Springer}, \bibinfo{address}{New York},
  \bibinfo{year}{1996}), \bibinfo{edition}{2nd} ed.

\bibitem[{\citenamefont{Novotn\'y}(2002)}]{Novotny2002a}
\bibinfo{author}{\bibfnamefont{T.}~\bibnamefont{Novotn\'y}},
  \bibinfo{journal}{Europhys. Lett.} \textbf{\bibinfo{volume}{59}},
  \bibinfo{pages}{648} (\bibinfo{year}{2002}).

\bibitem[{\citenamefont{Kohler et~al.}(2005)\citenamefont{Kohler, Lehmann, and
  H\"anggi}}]{Kohler2005a}
\bibinfo{author}{\bibfnamefont{S.}~\bibnamefont{Kohler}},
  \bibinfo{author}{\bibfnamefont{J.}~\bibnamefont{Lehmann}}, \bibnamefont{and}
  \bibinfo{author}{\bibfnamefont{P.}~\bibnamefont{H\"anggi}},
  \bibinfo{journal}{Phys. Rep.} \textbf{\bibinfo{volume}{406}},
  \bibinfo{pages}{379} (\bibinfo{year}{2005}).

\bibitem[{\citenamefont{Lax}(1963)}]{Lax1963a}
\bibinfo{author}{\bibfnamefont{M.}~\bibnamefont{Lax}}, \bibinfo{journal}{Phys.
  Rev.} \textbf{\bibinfo{volume}{129}}, \bibinfo{pages}{2342}
  (\bibinfo{year}{1963}).

\bibitem[{\citenamefont{van Kampen}(1992)}]{vanKampen1992a}
\bibinfo{author}{\bibfnamefont{N.~G.} \bibnamefont{van Kampen}},
  \emph{\bibinfo{title}{Stochastic processes in physics and chemistry}}
  (\bibinfo{publisher}{North-Holland}, \bibinfo{address}{Amsterdam},
  \bibinfo{year}{1992}).

\bibitem[{\citenamefont{Petersson et~al.}(2009)\citenamefont{Petersson, Smith,
  Anderson, Atkinson, Jones, and Ritchie}}]{Petersson2009a}
\bibinfo{author}{\bibfnamefont{K.~D.} \bibnamefont{Petersson}},
  \bibinfo{author}{\bibfnamefont{C.~G.} \bibnamefont{Smith}},
  \bibinfo{author}{\bibfnamefont{D.}~\bibnamefont{Anderson}},
  \bibinfo{author}{\bibfnamefont{P.}~\bibnamefont{Atkinson}},
  \bibinfo{author}{\bibfnamefont{G.~A.~C.} \bibnamefont{Jones}},
  \bibnamefont{and} \bibinfo{author}{\bibfnamefont{D.~A.}
  \bibnamefont{Ritchie}}, \bibinfo{journal}{Phys. Rev. Lett.}
  \textbf{\bibinfo{volume}{103}}, \bibinfo{pages}{016805}
  (\bibinfo{year}{2009}).

\end{thebibliography}
\end{document}